\documentclass[12pt]{iopart}
\usepackage{graphicx}  
\begin{document}

\title[An Analytic Model of Superclusters]{An Analytic Model of Superclusters}

\author{Masamune Oguri\dag\ddag, Keitaro Takahashi\ddag, 
Kiyotomo Ichiki\S$\|$ and Hiroshi Ohno\S\P
}

\address{\dag Princeton University Observatory, Peyton Hall,
Princeton, NJ 08544, USA.}
\address{\ddag Department of Physics, University of
Tokyo, Hongo 7-3-1, Bunkyo-ku, Tokyo 113-0033, Japan}
\address{\S Division of Theoretical Astrophysics, National
Astronomical Observatory, 2-21-1 Osawa, Mitaka, Tokyo 181-8588, Japan}
\address{$\|$ Department of Astronomy, University of Tokyo, Hongo 7-3-1,
Bunkyo-ku, Tokyo 113-0033, Japan} 
\address{\P Institute for Cosmic Ray Research, University of Tokyo,
5-1-5 Kashiwa, Kashiwa City, Chiba 277-8582, Japan}

\begin{abstract}
We construct an analytic model for the mass distributions of
superclusters and clusters in each supercluster. Our model is a
modification of the Press-Schechter theory, and defines superclusters 
as the regions that have some overdensity smaller than those for
usual virialized objects. We compare the mass functions with 
a catalog of superclusters in the Sloan Digital Sky Survey 
Early Data Release and found that they are in reasonable agreement with
each other.
\end{abstract}



\maketitle

\section{Introduction}

It is now known that our universe is hierarchical. The largest
virialized objects in the universe are clusters that are made of
hundreds of galaxies. Clusters have extensively been used to
test structure formation models in the universe, through their
abundances and correlations\cite{bahcall83,bahcall92}.

Superclusters are defined as an ensemble of clusters. They are still 
assembling, and are even larger structure than clusters;
superclusters have extents of $\sim 100{\rm Mpc}$ or even larger.
Thus superclusters offer us information on late evolution of universe, 
and also on the transition from linear to non-linear regime. 
Examples of the superclusters include A901/902\cite{gray02}, 
MS0302+17\cite{fabricant94,gavazzi04}, Cl1604+43\cite{lubin00,gal04}, 
the Sculptor supercluster\cite{schuecker91}, 
and the Shapley supercluster\cite{quintana95}. 
Detailed investigations have shown that these superclusters are real
physical systems where clusters are gravitationally assembling.
Detailed studies of these superclusters from galaxy distributions and
weak lensing also showed that the mass distribution of superclusters 
is in good agreement with the distribution of early-type galaxies 
in superclusters. The physical state of gas in superclusters is still
unknown, but it is worth mentioning that soft X-ray has been detected in
the Sculptor supercluster\cite{zappacosta04}.

Although most the studies have been limited to those of individual
superclusters so far, the emergence of wide-field surveys/telescopes,
such as the Sloan Digital Sky Survey (SDSS) and the Subaru Suprime-cam,
begins to make it possible to do {\it statistical} studies of 
superclusters. Nevertheless, there is no analytic model of superclusters 
that predicts, e.g., the mass distribution. In this paper, we construct
such an analytic model of superclusters. Basically, our model is a
modification of the Press-Schechter theory\cite{press74}, and predicts
not only mass distributions of superclusters but also number
distribution of clusters in each supercluster. We also compare our model
with superclusters in the SDSS extracted by Einasto {\it et
al.}\cite{einasto03}. Throughout the paper we assume a flat universe 
$\Omega_M+\Omega_\Lambda=1$, and the Hubble constant $h\equiv
H_0/(100{\rm km\,s^{-1}Mpc^{-1}})=0.7$.

\section{An Analytic Model}

In this section, we construct the mass function of superclusters and
clusters in each supercluster using the spherical model\cite{press74}.
We regard superclusters as overdense regions that are on the course of
the spherical collapse. Then superclusters can be defined as the regions 
that have the overdensity $\Delta_{\rm sc}\equiv \rho_{\rm
sc}/\bar{\rho}$, where $\rho_{\rm sc}$ is the average density within
superclusters and $\bar{\rho}$ is the mean density in the universe.
Although superclusters are usually characterized by filamentary structure 
that is quite different from simple spherical overdensity, we consider
this model so as to be analytically tractable. The assumption, however, 
may not be so crude given the similar situation that the Press-Schechter
theory has achieved reasonable success despite its simplicity.

\subsection{Spherical Model in Lambda-dominated Universes}

\begin{figure}[tb]
\begin{center}
\includegraphics[width=0.5\textwidth]{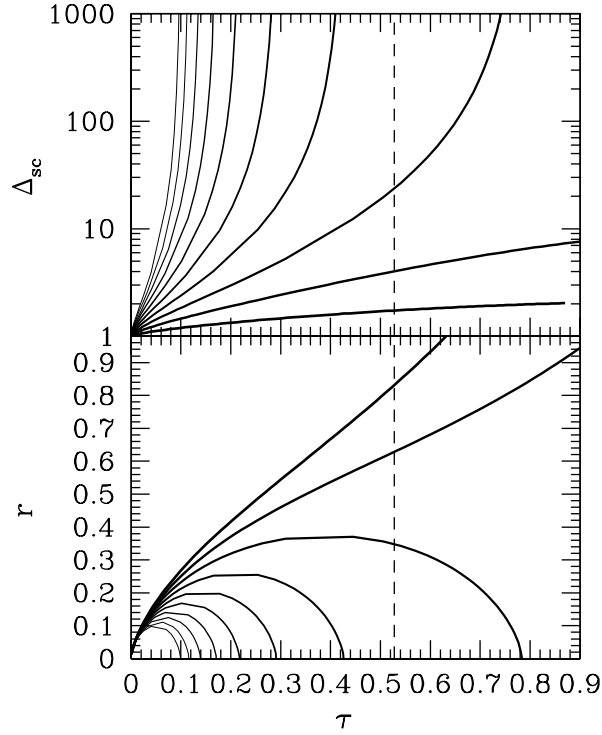}
\end{center}
\caption{Sizes of overdense regions ($r$) and corresponding nonlinear
 overdensities ($\Delta_{\rm sc}$) as a function of dimensionless cosmic
 time $\tau$. Different lines mean different values of $\kappa$: from
 $\kappa=1$ ({\it Dark}) to $10$ ({\it Light}). The vertical dashed line
 indicates the value of $\tau$ corresponding to $a=1$. In this plot, we
 adopt $\Omega_M=0.3$. \label{fig:sc}} 
\end{figure}

\begin{figure}[tb]
\begin{center}
\includegraphics[width=0.5\textwidth]{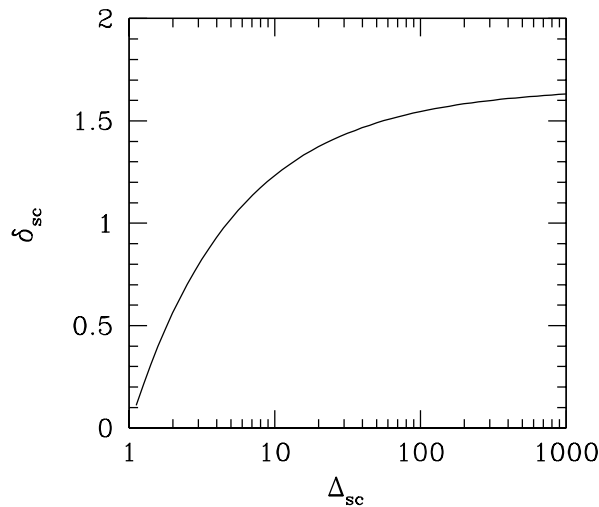}
\end{center}
\caption{The relation of linear (eq. [\ref{dlin}]) and nonlinear
 (eq. [\ref{dnl}]) overdensities at $a=1$. 
Here we adopt $\Omega_M=0.3$. 
\label{fig:ov}} 
\end{figure}

Consider a spherical overdensity with mass $M$ and radius $R$. 
Equation of motion of the overdensity is simply given by
\begin{equation}
 \frac{d^2R}{dt^2}=-\frac{GM}{R^2}+\Omega_\Lambda H_0^2R.
\end{equation}
By integrating this equation, we obtain
\begin{equation}
 -E=\frac{1}{2}\left(\frac{dR}{dt}\right)^2-\frac{GM}{R}-\frac{1}{2}\Omega_\Lambda H_0^2R^2,
\label{ene}
\end{equation}
here $E>0$ for a growing perturbation. We define
\begin{eqnarray}
 r &\equiv& \frac{R}{R_0},\\
 R_0&\equiv&\left(\frac{2GM}{H_0^2\Omega_M}\right)^{1/3},\\
 \kappa&\equiv&\frac{2E}{H_0^2\Omega_MR_0^2},\\
 \tau&\equiv&\sqrt{\Omega_M} H_0t,\\
 \omega &\equiv&\frac{1}{\Omega_M}-1.
\end{eqnarray}
Then equation (\ref{ene}) reduces to
\begin{equation}
\left(\frac{dr}{d\tau}\right)^2=\frac{1}{r}-\kappa+\omega r^2.
\label{col}
\end{equation}
On the other hand, the cosmological time is related with the scale
factor as
\begin{equation}
 \tau=\frac{1}{3}\omega^{-1/2}{\rm arccosh}\left(1+2\omega a^3\right).
\label{age}
\end{equation}
From equation (\ref{col}), we can derive the motion of spherical
overdensity as a function of $\tau$. Figure \ref{fig:sc} shows the
evolution of the size of overdense region for various values of $\kappa$.
It is shown that the corresponding overdensity $\Delta_{\rm sc}$
increases monotonically as time evolves. Since the dimensionless time
$\tau$ is related with the scale factor $a$ as equation (\ref{age}),
at fixed time $a$ and for fixed  $\Delta_{\rm sc}$ we can determine
the value of $\kappa$ uniquely by solving equations (\ref{col}) and
(\ref{age}). Then the linear and nonlinear overdensity is easily derived:
\begin{eqnarray}
 \delta_{\rm sc}&=&\frac{3\kappa}{5}D(a),\label{dlin}\\
 \Delta_{\rm sc}&=&\frac{a^3}{R^3},\label{dnl}
\end{eqnarray}
where $D(a)$ is the linear growth rate normalized so that $D(a)=a$ 
at $a\ll 1$. From this, arbitrary $\Delta_{\rm sc}$ is converted to
the corresponding linear overdensity $\delta_{\rm sc}$ that is an
essential quantity in computing the mass function. In Figure
\ref{fig:ov}, we plot an example of this relationship. It shows
one-to-one correspondence between $\Delta_{\rm sc}$ and $\delta_{\rm
sc}$ . 

\subsection{Mass Function of Superclusters and Clusters}

To compute the mass function of superclusters, we adopt the
Press-Schechter formalism\cite{press74}. From the linear overdensity
$\delta_{\rm sc}$ (eq. [\ref{dlin}]), the mass function can be computed as
\begin{equation}
 \frac{dn_{\rm sc}}{dM}=\sqrt{\frac{2}{\pi}}\frac{\rho_0}{M}\frac{\delta_{\rm sc}(z)}{\sigma_M^2}\left|\frac{d\sigma_M}{dM}\right|\exp\left(-\frac{\delta_{\rm sc}^2(z)}{2\sigma_M^2}\right),
\label{mf}
\end{equation}
where $\delta_{\rm sc}(z)\equiv\delta_{\rm sc}\{D(z=0)/D(z)\}$, and
$\sigma_M$ is the (linear) mass variance at $z=0$. 

The conditional mass function of clusters within superclusters, that are
already virialized, can be estimated in analog with the extended
Press-Schechter formalism\cite{bower91,bond91,lacey93}. The number of
clusters with mass $M$  that are in a supercluster with mass $M_{\rm
sc}$ is given by  
\begin{equation}
 \frac{dn}{dM}(M|M_{\rm sc};z)=\frac{1}{\sqrt{2\pi}}\frac{M_{\rm sc}}{M}\frac{\delta_{\rm c}(z)-\delta_{\rm sc}(z)}{\left(\sigma^2_M-\sigma^2_{M_{\rm sc}}\right)^{3/2}}\left|\frac{d\sigma^2_M}{dM}\right|\exp\left[-\frac{\left(\delta_{\rm c}(z)-\delta_{\rm sc}(z)\right)^2}{2\left(\sigma^2_M-\sigma^2_{M_{\rm sc}}\right)}\right],
\label{cmf}
\end{equation}
where $\delta_{\rm c}(z)\approx 1.68\{D(z=0)/D(z)\}$ is the
critical overdensity in the standard Press-Schechter formalism.
Although the distribution of clusters in each supercluster might 
be affected by physical processes such as global tides and dynamical
frictions\cite{oguri04}, we adopt this form for simplicity. 
\section{Comparison with Observations}

\begin{figure}[tb]
\begin{center}
\includegraphics[width=0.5\textwidth]{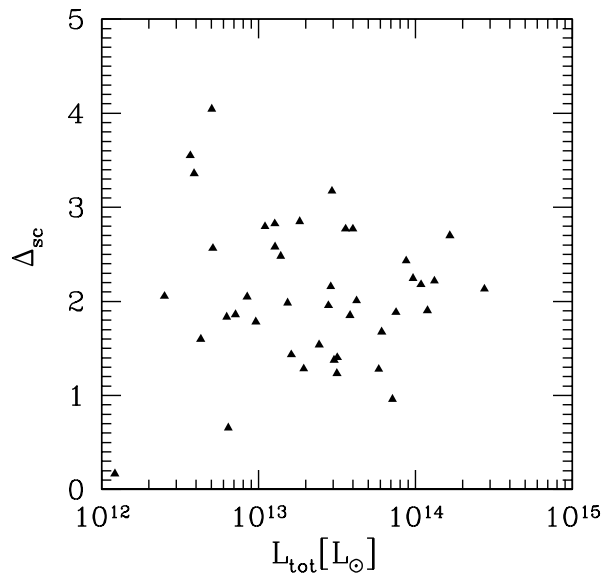}
\end{center}
\caption{Overdensities versus total luminosities of superclusters in the
 SDSS EDR\cite{einasto03}. We assumed the average $r$-band luminosity
 density  $\bar{\rho}_{\rm L}=2.6\times10^8hL_\odot{\rm Mpc^{-3}}$\cite{blanton01}
 to convert into $\Delta_{\rm sc}$. 
\label{fig:zdel}} 
\end{figure}

In this section, we compare our analytic model with the results of
Einasto {\it et al.}\cite{einasto03} who made a catalog of
superclusters in the SDSS Early Data Release (EDR)\cite{stoughton02}.
They extracted superclusters from the density field of galaxies with a
smoothing length 10$h^{-1}{\rm Mpc}$, and also clusters from the density
field  with a smoothing length 0.8$h^{-1}{\rm Mpc}$. The superclusters
and clusters are defined as the regions divided by the threshold density
($\delta=1.8$). Since the SDSS EDR data consists of two slices, the
density filed is calculated in 2 dimensions only. Thus the total
($r$-band) luminosity of a supercluster is derived as
\begin{equation}
L_{\rm tot}=\frac{D}{D_{\rm d}}L_{\rm obs}, 
\label{corr}
\end{equation}
where $D_{\rm d}$ is the thickness of the slice, $D$ is the average
diameter of the supercluster, and $L_{\rm obs}$ is the observed
luminosity with the selection function of galaxies corrected. With this
way 39 superclusters were identified (see Table 3 of \cite{einasto03}).

Since the above definition of superclusters is slightly different from 
theoretical one, we need to check what the observed superclusters
correspond to in our model. We have defined superclusters as the regions 
that have the overdensity $\Delta_{\rm sc}$, so we compute $\Delta_{\rm
sc}$ for each observed supercluster. Figure \ref{fig:zdel} shows
$\Delta_{\rm sc}$ for the SDSS EDR superclusters. Here we calculated
$\Delta_{\rm sc}$ as 
\begin{equation}
\Delta_{\rm sc}=\frac{L_{\rm tot}/\{(4/3)\pi (D/2)^3\}}{\bar{\rho}_{\rm L}}, 
\end{equation}
where $\bar{\rho}_{\rm L}=2.6\times10^8hL_\odot{\rm Mpc^{-3}}$ is the average
$r$-band luminosity density\cite{blanton01}. This coincides with
equation (\ref{dnl}) if the mass-to-light ratio of superclusters is the
same as that of the universe. As seen in the Figure, the derived value
of $\Delta_{\rm sc}$ is scattered at around $2$ ($\Delta_{\rm sc} = 2.1
\pm 0.8$).  Thus we adopt 
$\Delta_{\rm sc}=2$ as a fiducial value, though we also examine the
cases with $\Delta_{\rm sc}\neq 2$ to see how the uncertainty of
$\Delta_{\rm sc}$ would affect our results.

\subsection{Mass Function of Superclusters}

\begin{figure}[tb]
\begin{center}
\includegraphics[width=0.7\textwidth]{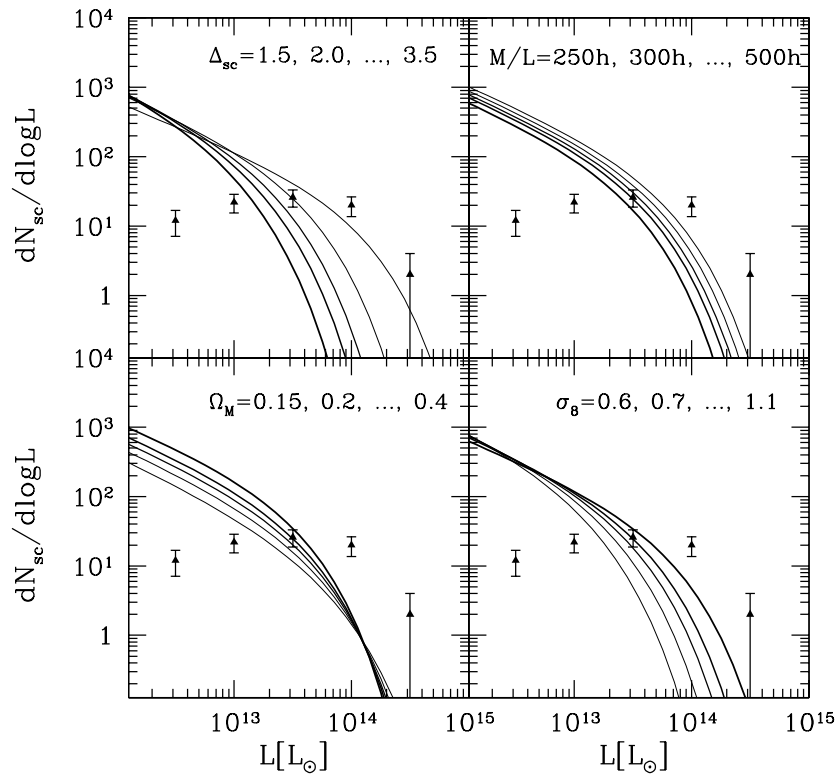}
\end{center}
\caption{Comparison of luminosity functions of superclusters. The
 observational data ({\it filled triangles with errorbars})
 are taken from \cite{einasto03}. For the errorbars, only statistical
 errors are included.  We adopt the following fiducial
 model in computing theoretical predictions: $\Delta_{\rm sc}=2.0$,
 $M/L=400hM_\odot/L_\odot$, $\Omega_M=0.3$, and $\sigma_8=0.9$. For each
 panel, we changed one of these parameters while the other parameters
 are fixed. In all panels, the darker lines mean the larger values.
\label{fig:lf}} 
\end{figure}

First, we compare the abundance of superclusters in the SDSS EDR field.
To do so, the volume of the field where superclusters were identified 
is needed. We simply adopt the sum of volumes of North and South slices
of the EDR data\cite{einasto03}. However, this may
underestimate the volume because of the correction outside the slices
(eq. [\ref{corr}]). To convert mass functions of superclusters to
luminosity functions, the mass-to-light ratio of superclusters. Again,
we assume the fiducial model $M/L=400hM_\odot/L_\odot$, and estimate the
sensitivity by changing $M/L$ from $250hM_\odot/L_\odot$ to
$500hM_\odot/L_\odot$. 

Figure \ref{fig:lf} plots luminosity functions of superclusters in
theory and observations. First, the abundance of massive superclusters
is sensitive to $\Delta_{\rm sc}$, thus the selection of superclusters
is quite important in comparing the abundance. Next, the abundance of
massive superclusters is also very sensitive to $\sigma_8$. 
Therefore, surveys of massive superclusters can be a good probe of 
$\sigma_8$. Finally, the observed number of superclusters shows
reasonable agreement with the theory, particularly when we take account
of several uncertainties including that of the survey volume. At smaller
luminosities the difference seems large, but this may be ascribed to the
incompleteness of observations for such small objects.

\subsection{Number of Clusters in Superclusters}

Einasto {\it et al.} also estimated the number of clusters in each
supercluster. They noted that there exists a well-defined lower limit of
cluster luminosities that is a function of distance to clusters (denoted
by $d$), particularly for those at larger distances (see Figure 6 of
\cite{einasto03}). Thus we divide the supercluster sample by the
distance $d$. For each subsample, we assume the limiting luminosity
shown in Figure 6 of \cite{einasto03} and calculate the number of
clusters in a given supercluster by integrating equation (\ref{cmf}). 
We also convert the observed number of clusters in the EDR field to the {\it
total} number in the similar way as equation (\ref{corr}).

\begin{figure}[tb]
\begin{center}
\includegraphics[width=0.7\textwidth]{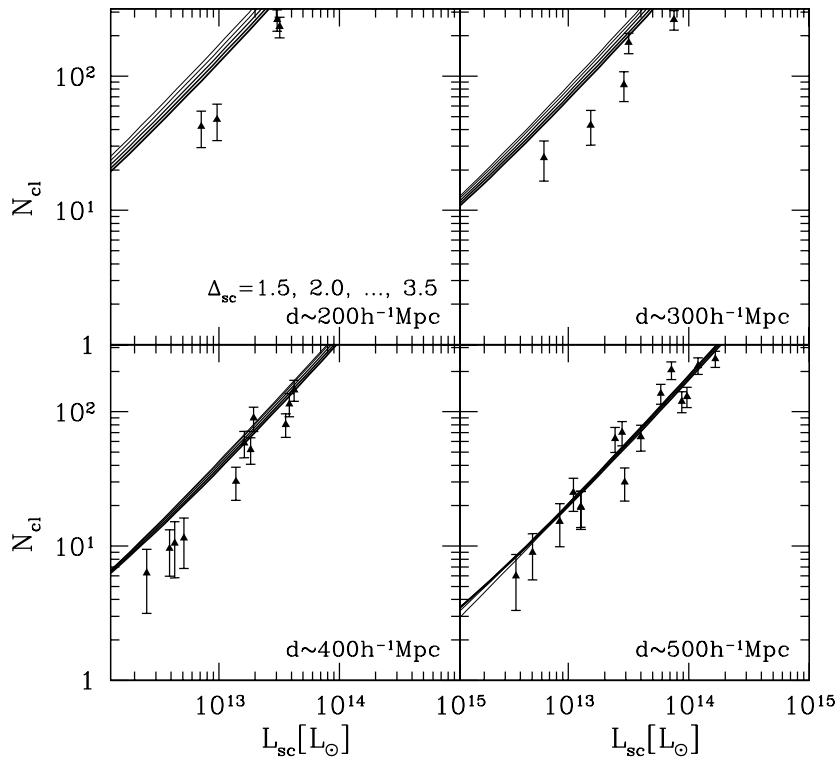}
\end{center}
\caption{Comparison of the numbers of clusters in superclusters. 
 The observational data ({\it filled triangles})
 are taken from \cite{einasto03}. For the errorbars, only statistical
 errors are included.  
 We consider four subsamples of superclusters divided by the distances
 to superclusters; $d\sim 200h^{-1}{\rm Mpc}$ 
 ($150h^{-1}{\rm Mpc}<d<250h^{-1}{\rm Mpc}$; {\it upper left}), 
 $d\sim 300h^{-1}{\rm Mpc}$ 
 ($250h^{-1}{\rm Mpc}<d<350h^{-1}{\rm Mpc}$; {\it upper right}), 
 $d\sim 400h^{-1}{\rm Mpc}$ 
 ($350h^{-1}{\rm Mpc}<d<450h^{-1}{\rm Mpc}$; {\it lower left}), and
 $d\sim 500h^{-1}{\rm Mpc}$ 
 ($450h^{-1}{\rm Mpc}<d<550h^{-1}{\rm Mpc}$; {\it lower right}).
 As limiting masses of clusters, we adopt ``set 1'' in Figure 6 of
 \cite{einasto03}.  Different lines mean different values of
 $\Delta_{\rm sc}$. The other parameters are fixed to the fiducial
 values (see the caption of Figure \ref{fig:lf}).
\label{fig:cl_dis}} 
\end{figure}

\begin{figure}[tb]
\begin{center}
\includegraphics[width=0.7\textwidth]{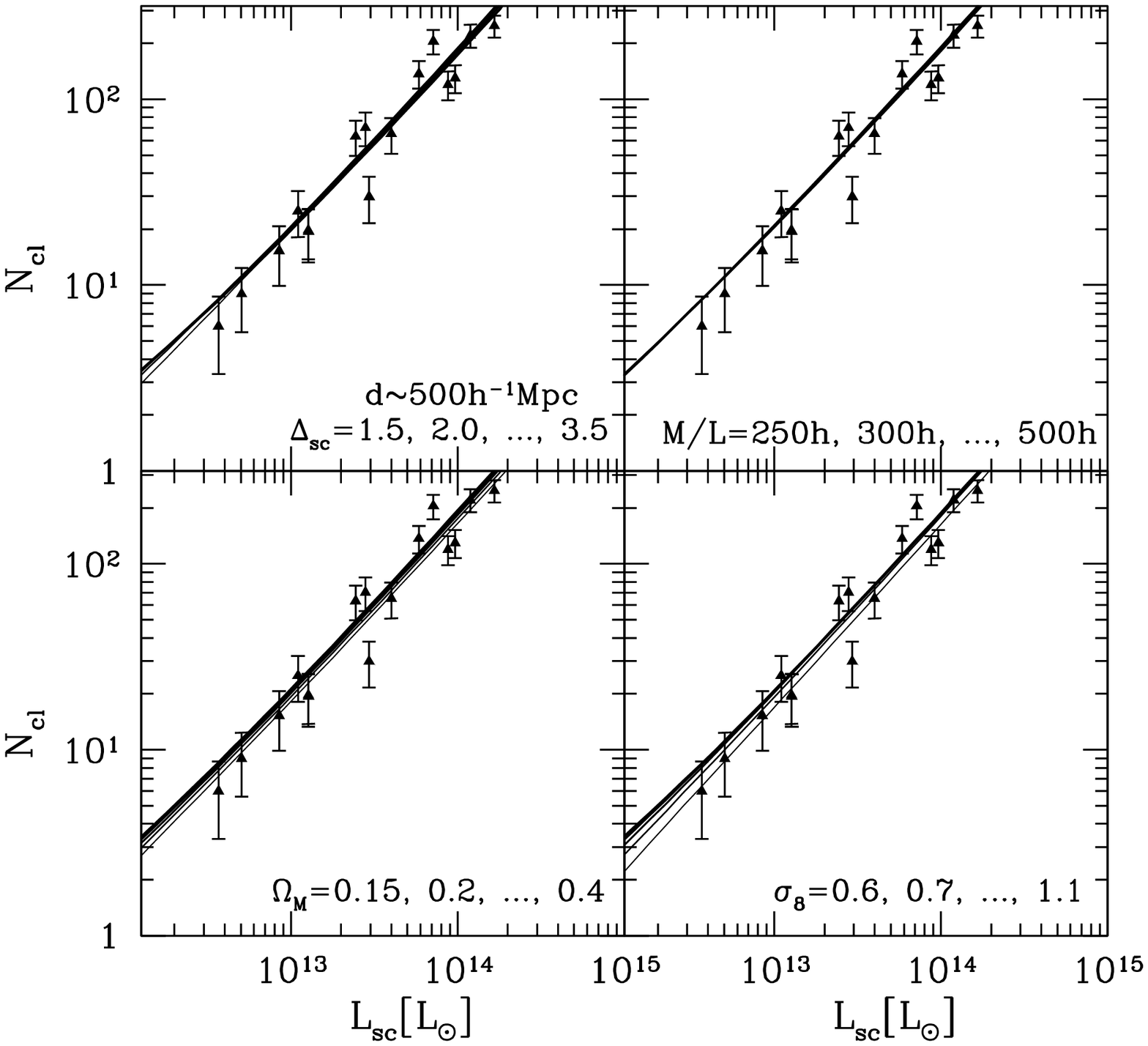}
\end{center}
\caption{Comparison of the numbers of clusters in superclusters at
 $d\sim 500h^{-1}{\rm Mpc}$.  The observational data ({\it filled
 triangles}) are taken from \cite{einasto03}. For the errorbars, only
 statistical errors are included.  For each panel, we
 changed one parameters while the other parameters are fixed to the
 fiducial values (see the caption of Figure \ref{fig:lf}). In all panels,
 the darker lines mean the larger values.
\label{fig:cl_par}} 
\end{figure}

In Figure \ref{fig:cl_dis}, we plot the number of clusters within
superclusters for different distances $d$. We find that theoretical
curves and observations show good agreement with each other. It seems
that the model overpredicts the number of clusters, but this is perhaps
because the assumption of the existence of lower limit of cluster
luminosities becomes inaccurate for lower $d$ (see Figure 6
of \cite{einasto03}). Next we examine the dependence of several parameters
on the number of clusters, which is shown in Figure \ref{fig:cl_par}.
The Figure indicates that the number of clusters in superclusters hardly
depends on these parameters, though that was already clear from the
expression of the conditional mass function shown in equation
(\ref{cmf}). Thus the model predictions are quite
robust, and the good agreement means that our model works  well.
 
\section{Summary and Discussion}

In this paper, we have constructed an analytic model of superclusters. Our
model is based on the Press-Schechter theory, and defines superclusters
as overdense regions that are on the course of the spherical collapse. 
Our model allows us to compute the mass function of superclusters, and
also the number distributions of clusters in each supercluster.

To test the model, we have compared the model with the supercluster
catalog in the SDSS EDR data. We have shown that the model shows in
reasonable agreement with the observation. We have also examined 
parameter dependences of the mass function of superclusters, and found
that the mass function of massive superclusters is sensitive to
$\sigma_8$. Thus superclusters will be a useful probe of $\sigma_8$ if
superclusters are selected more rigorously. On the other hand, the
number of clusters in superclusters shows only weak dependences on
parameters. 

Our model would be useful not only in constraining cosmological
parameters from the abundance of superclusters but also in many
applications. One of such applications is hot gas in superclusters. 
Recently it has been claimed that soft X-ray\cite{zappacosta04} and 
Sunyaev-Zel'dovich effects\cite{myers04} in the supercluster regions 
may be detected, and this might suggest the existence of relatively 
hot gas in superclusters. If this is true, the statistical significance
of the signals that are caused by the hot gas can be computed by using 
our model, once the we measure the temperature and density structure of
hot gas. Other application is gravitational lensing; gravitational
lensing measures the total mass along the line of sight, thus
a supercluster outside a cluster may affect, e.g., the mass estimate of
the cluster. Our model is useful since it allows us to estimate the
distribution of the host superclusters for a given cluster.
In practice, we found that the effect is negligible when we consider
superclusters with $\Delta_{\rm sc}\sim 2$; assuming a homogeneous
sphere, convergence of a supercluster with $M=10^{16}h^{-1}M_\odot$ can be
estimated as 
\begin{equation}
\kappa=\frac{\Sigma}{\Sigma_{\rm crit}}\sim \frac{9\times 10^{12}(M/10^{16}h^{-1}M_\odot)^{1/3}(\Delta_{\rm
 sc}/2)^{2/3}hM_\odot{\rm Mpc^{-2}}}{4\times 10^{15}hM_\odot{\rm
 Mpc^{-2}}}\sim 2\times10^{-3},
\end{equation}
where we assumed a source at $z\sim 1$ and a supercluster at $z\sim 0.3$.
However, it is of interest to consider the effect more carefully,
because the effect is dependent on the density profile of superclusters
as well as the definition of superclusters, $\Delta_{\rm sc}$. In
particular, superclusters could have much more significant effect if 
superclusters have centrally concentrated density distributions. 

Although our model looks successful at this moment, definitely we need
more tests. The more direct way to test the model is to compare it with
numerical simulations, since we know the initial condition and
cosmological parameters in numerical simulations. Of course, comparison
with observations is also of great importance in order to test our
understanding of the universe. Now the SDSS data have been becoming
public\cite{abazajian03,abazajian04}, and the quite large survey
region of the SDSS would allow us to make three-dimensional catalog of
superclusters. The follow-up work is currently in progress, and the
results will be presented elsewhere.  

\section*{Acknowledgements}
We thank Issha Kayo for useful discussions. MO, KT, and KI 
are supported by JSPS through JSPS Research Fellowship
for Young Scientists.  

\section*{References}


\end{document}